\documentclass[amssymb,useAMS,prd,aps,twocolumn,superscriptaddress]{revtex4}
\usepackage{pslatex}
\usepackage{graphicx}
\usepackage{psfrag}

\def\refe@jnl#1{{#1}}
\def\aj{\refe@jnl{Astron.~J.}}
\def\araa{\refe@jnl{Annu.~Rev.~Astron.~Astrophys.}}
\def\apj{\refe@jnl{Astrophys.~J.}}
\def\apjl{\refe@jnl{Astrophys.~J.~Lett.}}
\def\aap{\refe@jnl{Astron.~Astrophys.}}
\def\mnras{\refe@jnl{Mon.~Not.~R.~Astron.~Soc.}}
\def\prd{\refe@jnl{Phys.~Rev.~D}}
\def\fcp{\refe@jnl{Fund.~Cos.~Phys.}}
\def\physrep{\refe@jnl{Phys.~Rep.}}
\def\physlett{\refe@jnl{Phys.~Lett.}}

\def\invisible#1{  }

\begin{document}

\title{Are small neutrino masses unveiling the missing
mass problem of the Universe?}

\author{C\'eline B\oe hm}

\affiliation{LAPTH, UMR 5108, 9 chemin de Bellevue - BP 110,
  74941 Annecy-Le-Vieux, France.}

\author{Yasaman Farzan}
\affiliation{Institute for Studies in Theoretical Physics and
  Mathematics (IPM), P.O. Box 19395-5531, Tehran, Iran}

\author{Thomas Hambye}
\affiliation{Instituto de F\'isica Te\'orica, Universidad Aut\'onoma
de Madrid, Cantoblanco, Spain}

\author{Sergio Palomares-Ruiz}
\affiliation{Institute for Particle Physics Phenomenology,
University
  of Durham, Durham DH1 3LE, United Kingdom}

\author{Silvia Pascoli}
\affiliation{Institute for Particle Physics Phenomenology,
University
  of Durham, Durham DH1 3LE, United Kingdom}

\date{today}

\begin{abstract}
We present a scenario in which a remarkably simple relation linking
dark matter properties and neutrino masses naturally emerges. This
framework points towards a low energy theory where the neutrino mass
originates from the existence of a light scalar dark matter particle
in the MeV mass range. A very surprising aspect of this scenario is
that the required MeV dark matter is one of the favored candidates
to explain the mysterious emission of 511 keV photons from the
center of our galaxy. A possible interpretation of these findings is
that dark matter is the stepping stone of a theory beyond the
standard model instead of being an embarrassing relic whose energy
density must be accounted for in any successful model building.\\
Lapth-1169/06; IPM/P-2006/077; IPPP/06/84; DCPT/06/168.
\end{abstract}

 \maketitle

\section{Introduction}
The discovery of non-zero neutrino masses in neutrino oscillation
experiments~\cite{nuosci} and the increasing evidence for about 23
$\%$ of the content of the Universe being in the form of dark
matter~\cite{DMhint} are the two main indications for physics beyond
the Standard Model. These two issues, the origin of neutrino masses
and the nature of dark matter, have been long standing problems in
particle physics. Yet, in general, they are considered as two
different topics and current explanations rely on completely
different mechanisms, involving unrelated particles and scales.
Although there have been some proposals to establish a
link~\cite{nuDM1,nuDM2,nuDM3,Ma:2006km,Ma06a,Ma06b,Ma06c,nuDM4,nuDM5},
a simple and natural picture in which the neutrino mass scale, the
dark matter properties and dark matter abundance would be
quantitatively related is still missing. In particular, to the best
of our knowledge, there is no model in the literature which uniquely
determines the dark matter scale and predicts, at the same time, a
direct connection between the smallness of neutrino masses and the
observed dark matter relic density.

 In this letter, we present a scenario where such
a prediction exists and therefore establish
a quantitative link between these two fields. Our framework is
strongly inspired by the class of models  proposed independently
in Ref.~\cite{Palomares-Ruiz:2005vf} to explain the signal
observed in the LSND neutrino experiment~\cite{LSND}, in
Refs.~\cite{bens,bf} to illustrate that annihilating dark matter
can be as light as a few MeV and in Ref.~\cite{511,Ascasibar} to
explain the mysterious presence of low energy positrons in the
center of our galaxy \cite{SPI}. It is based on the following
Lagrangian:
\begin{equation}
\label{lagrangian} {\cal{L}}_{I} \supset g \ \phi \ \bar{N} \ \nu_L
\end{equation}
where $g$ is a coupling constant, $N$ is a Majorana neutrino (with a
mass $m_N$), $\phi$ is a neutral scalar (singlet of $SU(2)_L\times
U(1)$) which plays the role of dark matter (hereafter referred to as
the \textit{SLIM} particle for Scalar as LIght as MeV), and $\nu_L$
is the standard left-handed neutrino. Since the mass of the particle
$N$ is of Majorana type, lepton number is not conserved. As one can
notice, the Lagrangian above contains only one interaction term.
Since it breaks the electroweak symmetry for the case that we
detail, it has to be regarded as a low energy effective Lagrangian.

Here we show that, with such a Lagrangian, a remarkably simple
relationship between the dark matter cross section and the neutrino
mass scale naturally emerges. Moreover the requirement of sub-eV
neutrino masses, as imposed by experimental constraints, points
towards light dark matter particles (with a mass of a few MeV). Our
expression therefore suggests that the issues regarding the dark
matter and neutrino masses are not only closely related but they
also share the same low energy origin. A possible interpretation of
these findings is that dark matter is fundamental. It may be the
first step towards a low energy theory beyond the standard model.

\section{Linking dark matter and neutrino mass}
In the Lagrangian given in Eq.~(\ref{lagrangian}), $\phi$ is a
scalar particle which may either be real or complex, and $N$ is a
heavy neutrino with a Majorana mass $m_N > m_{\phi}$. The particle
$\phi$ is stable (it cannot decay into $N$) and constitutes our dark
matter candidate. In contrast $N$  decays into $\phi$ and $\nu_L$
with a decay rate $\Gamma_N = g^2 m_N^2/(16 \pi E_N)$.

\begin{figure}[h]
  \includegraphics[width=5.5cm]{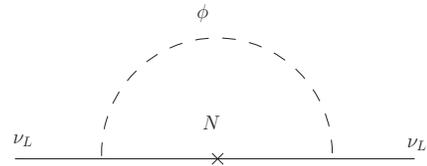}
  \caption{In the scenario discussed here, left-handed neutrinos
  acquire a very small
  mass term  due to their interactions with
  a SLIM and a $N$ particle. This mass term
  is of Majorana type owing to the Majorana nature of $N$.} \label{neutrino1}
\end{figure}

One consequence of the Lagrangian given in Eq.~\ref{lagrangian}  is
that left-handed neutrinos acquire a mass term $m_{\nu_L}$ via the
one-loop correction depicted in
Fig.\ref{neutrino1}~\cite{Ma:1998dn}. This mechanism is the same as
in Refs.~\cite{Ma:1998dn,Ma:2006km,Ma06a,Ma06b,Ma06c} except that,
in our scenario, $\phi$ is a singlet under the electroweak symmetry.
Like in Refs.~\cite{Ma:2006km,Ma06a,Ma06b,Ma06c}, we assume that
$\phi$ does not have a vacuum expectation value, so
Eq.~\ref{lagrangian} does not induce any tree level contribution to
the left-handed neutrino mass which could dominate over the
contribution discussed in this letter.

In this scenario, light neutrinos $\nu_L$ are predicted to be
Majorana particles. This prediction is important because it can be
tested in neutrinoless double beta decay
experiments~\cite{doublebeta1,doublebeta2}.

A real scalar field $\phi$ gives a contribution to  $m_{\nu_L}$
which is given by:
\begin{equation}
m_{\nu_L} =  \frac{g^2}{16 \ \pi^2} \ m_N \
\left[\ln\left(\frac{\Lambda^2}{m_N^2}\right) -
\frac{m_{\phi}^2}{m_N^2-m_{\phi}^2}
  \ln\left(\frac{m_N^2}{m_{\phi}^2}\right) \right].
\label{neutrino2}
\end{equation}
 In this expression $g, \ m_N, m_{\phi}$ and $\Lambda$
 (the ultraviolet cut-off of the effective theory) are
 free parameters.

\begin{figure}[h]
  \includegraphics[width=8cm]{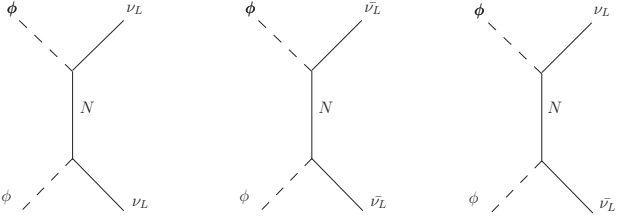}
  \caption{Here are the three diagrams corresponding to
  the main SLIM annihilation channels.
  These three diagrams are very similar
  to Fig.~\ref{neutrino1}. They involve the same
  couplings $g$ and rely on the exchange of the particle $N$.
  The important point is that the first diagram exactly
  corresponds to Fig.~\ref{neutrino1} if one joins the legs of the two
  SLIM particles in
  the initial state and  rotates the diagram.}
\label{annihil1}
\end{figure}

From Eq.~\ref{lagrangian}, one can draw the three diagrams shown in
Fig.~\ref{annihil1} which demonstrate that SLIM particles
 annihilate into two neutrinos (or two antineutrinos)
as well as neutrino-antineutrino pairs.

Owing to these annihilations, the SLIM number density decreases with
time. The rate at which the SLIM particles disappear is controlled
by the total SLIM pair annihilation cross section. The three
diagrams in Fig.~\ref{annihil1} are the only annihilation channels
available at tree-level. Hence the sum of these three contributions
sets the annihilation rate and therefore determines the SLIM relic
density.

The cross sections associated with the SLIM pair annihilations into
either neutrino or antineutrino pairs (see the first two diagrams of
Fig.~\ref{annihil1}), times $\rm{v}_r$ --the relative velocity of
the initial state particles--, are given by:
\begin{equation}
\langle \sigma(\phi \phi \to \nu_L \nu_L) {\rm{v}}_r \rangle
=\langle \sigma(\phi \phi \to \bar{\nu}_L \bar{\nu}_L) {\rm{v}}_r
\rangle
 \simeq \frac{g^4}{4 \pi}  \
\frac{m_{N}^2}{(m_{\phi}^2 + m_{N}^2)^2}, \label{crosssection}
\end{equation}
where the notation $\langle ... \rangle$ denotes the thermal average
of the quantity in the brackets.

In contrast, the cross section associated with the SLIM pair
annihilations into neutrino-antineutrino pairs (see last diagram of
Fig.~\ref{annihil1}) is suppressed by the ratio $m_{\nu_L}^2/m_N^2$
(or $p_{dm}^2/m_N^2$ in the case of complex particles, where
$p_{dm}$ is the dark matter momentum). This cross section is
therefore negligible with respect to the two others.

Hence $\langle\sigma \rm{v}_r\rangle$ (the total annihilation cross
section times $\rm{v}_r$) approximately corresponds to $ 2 \
\langle\sigma(\phi \phi \to \nu_L \nu_L) \rm{v}_r\rangle$. Any
relationship which involves this quantity is necessarily related to
the dark matter abundance.

The remarkable point is that, for $m_\phi<m_N \ll \Lambda\sim
m_{electroweak}$, the second term in Eq.~(\ref{neutrino2})
  can be  neglected. Then, using
Eq.~(\ref{crosssection}), Eq.~(\ref{neutrino2}) can be  very
simply rewritten as:
\begin{equation}
m_{\nu_L} \ \simeq   \sqrt{\frac{\langle\sigma v_r\rangle}{128  \
\pi^{3}}} \ m_{N}^2 (1+m_\phi^2/m_{N}^2)\
\ln\left(\frac{\Lambda^2}{m_{N}^2}\right). \label{mnu1}
\end{equation}
This relation shows that, in our scenario, the left-handed neutrino
masses and the dark matter abundance are very strongly related.

The simplicity of Eq.~(\ref{mnu1}) implies that one can make firm
predictions. The order of magnitude of $m_{\nu_L}$ can be obtained
from measurements in neutrino experiments; The value of $\langle
\sigma \rm{v}_r \rangle$ is set by the requirement that the SLIM
relic density must correspond to the measured dark matter abundance.
Thus, the only free parameters in Eq.~\ref{mnu1} are $m_N$,
$m_\phi/m_N$ and $\Lambda$. Since the dependence of Eq.~(\ref{mnu1})
on $\Lambda$ is only logarithmic and since varying the ratio
$m_\phi/m_N$ between 0 and 1 does not modify the result of
Eq.~(\ref{mnu1}) by more than a factor 2,  $m_N$ is the only
important free parameter.

In order to explain the observed dark matter abundance, the total
SLIM pair annihilation cross section must be (see
Refs.~\cite{leeweinberg,bens}) of the order of~:
\begin{equation}
\langle \sigma v_r \rangle \simeq  \ 10^{-26} \, \rm{cm^3/s}.
\label{annihilation}
\end{equation}
It is worth noticing that this value is, in first approximation,
independent of the dark matter mass and corresponds to a coupling
\begin{equation}
\label{gvalues} g \simeq 10^{-3} \ \sqrt{\frac{m_N}{10 \
\rm{MeV}}} \ \left(\frac{\langle \sigma v_r\rangle}{10^{-26}
\rm{cm^3/s}}\right)^{1/4}\left({1+m_\phi^2/m_N^2}\right)^{1/2}.
\end{equation}

If we now insert Eq.~\ref{annihilation} into Eq.~\ref{mnu1} and
take, for instance, $\Lambda\sim E_{electroweak} \sim \, 200 \,
\rm{GeV}$ and consider $0.01~{\rm eV} < m_\nu < \, 1~{\rm eV}$, we
obtain that $m_N$ can only vary within the range~:
\begin{equation}
\label{MN}
 {\mathcal O(1)} \ {\rm{MeV}} \, \lesssim m_{ N } \lesssim \ 10 \,
 \rm{MeV}.
\end{equation}
This range will be narrowed down by improving the bounds on neutrino
masses or, possibly, by directly measuring them. To be accurate, one
should take into account flavor effects, i.e. one should specify the
combination of neutrino flavors to which $N$ is coupled, keeping in
mind that at least a second (heavier) $N$ is necessary to lead to at
least two massive neutrinos. This would be done in a forthcoming
paper.

Since $m_{\phi}<m_N$, we can therefore conclude from Eq.~\ref{MN}
that $m_{\phi} < 10$~MeV. The exact value of $m_N$ depends on the
actual cut-off $\Lambda$ of the theory but, as already mentioned,
this dependence is only logarithmic. Note that the above range
implies that $N$ is an electroweak singlet or has very weak
couplings to the the Standard Model $Z$ boson.

Combining Eqs.~\ref{gvalues} and \ref{MN}, we conclude that
\begin{equation}
3 \times 10^{-4} \lesssim g \lesssim 10^{-3}. \label{greal}
\end{equation}

Let us now discuss  the case of a complex scalar field,
$\phi=(\phi_1+i\phi_2)/\sqrt{2}$ where   $\phi_1$ and $\phi_2$ are
real fields with masses $m_{\phi_1}$ and $m_{\phi_2}$. The various
equations obtained for real $\phi$ are modified but the overall
picture remains the same. In particular, Eq.~\ref{neutrino2} becomes
\begin{eqnarray}
 m_{\nu_L} =  \frac{g^2}{32 \ \pi^2} \ m_N \ \left[\frac{m_{\phi
       1}^2}{(m_N^2-m_{\phi1 }^2)} \ln(m_N^2/m_{\phi 1}^2)
\right. \nonumber
   \\
- \left. \frac{m_{\phi 2}^2}{(m_N^2-m_{\phi 2}^2)} \ln(m_N^2/m_{\phi
2}^2) \right], \label{complexmass}
\end{eqnarray}
and Eq.~\ref{mnu1} becomes
\begin{equation}
\label{mnu2} m_{\nu_L} \ = \ \sqrt{\frac{\langle \sigma
v_r\rangle}{128 \ \pi^{3}}} \ \left( m_{\phi_1}^2 \ln
\frac{m_N^2}{m_{\phi_1}^2} - m_{\phi_2}^2 \ln \frac{m_N^2}{m_{\phi
2}^2} \right) ~,
\end{equation}
where we have neglected the terms suppressed by
$m_{\phi_{1,2}}^2/m_N^2$. Note that the cut-off dependence drops out
in Eq.~\ref{complexmass}. In Eq.~\ref{mnu2}, the neutrino mass is
determined by the quantity $m_{\phi_1}^2-m_{\phi_2}^2$ while, in
Eq.~\ref{mnu1}, it was determined by $m_{N}^2$. Hence, instead of
Eq.~\ref{MN}, we now obtain:
\begin{equation}
\label{Mphisplit} ( 1 \, {\rm{MeV}})^2 \, \lesssim
|m_{\phi_1}^2-m_{\phi_2}^2| \lesssim \ ( 20 \, \rm{MeV})^2.
\end{equation}
For definiteness, we have assumed that the ratio $m_N/m_{\phi_1}$
was ranging from 10 to $10^5$. In Eq.~\ref{mnu2}, the mass $m_N$ is
a free parameter which can be much larger than the mass of the $Z$
boson. Hence, in the complex case (unlike the real case), the
particle $N$ can have electroweak couplings.

 If, for example, $m_{\phi_2} < m_{\phi_1}$ one expects that
 the unstable particle, $\phi_1$, decays into $\phi_2$ plus
 a pair of neutrino and
antineutrino.
 The $\phi_2$ particle, being stable,  would be our
dark matter candidate. Note also that if the mass splitting between
$m_{\phi_1}$ and $m_{\phi_2}$ is small, one has to take into account
the coannihilations between $\phi_1$ and $\phi_2$ for the
calculation of the dark matter relic density. This may slightly
change Eqs.~\ref{mnu2} and ~\ref{Mphisplit}.

In summary, if $\phi$ is a real field, the natural scale for the
dark matter mass is the MeV range or below. As discussed in Section
III, a dark matter mass much smaller than a few MeV poses some
conflict with observations. Thus a dark matter mass in the MeV range
is the preferred solution in the real case. If $\phi$ is a complex
field, a suitable scale is also the MeV range although
Eq.~\ref{Mphisplit} does not uniquely predict that the dark matter
mass must lie in the low energy range.


Obtaining the MeV scale is quite an amazing finding since this
corresponds to the dark matter mass range which is required to
explain the 511 keV emission line from the center of our galaxy
\cite{bens,bf,511}.

Note that if $N$ is mixed with light neutrinos and has a mass $m_N
\lesssim 1$~MeV, it might be responsible for the LSND
signal~\cite{Palomares-Ruiz:2005vf}.

\section{Constraints}

The scenario that we discussed in the present letter satisfies the
constraints from direct and indirect dark matter detection
experiments. It also satisfies cosmological constraints. Large scale
structure arguments force the SLIM particle to have a mass greater
than a few keV, which is consistent with the present scenario. SLIM
particles are also consistent with the constraints obtained in
Refs.~\cite{structure1,structure2}.

In supernovae, the strong interactions between the SLIM particles
and neutrinos would maintain them in equilibrium. However, owing to
the weakness of the interactions (if any) between the SLIMs and the
rest of the Standard Model particles, neutrinos are emitted at
approximately the same temperature as in the standard scenario
without SLIM interactions. Thus, considering the present
observational, as well as theoretical uncertainties, no bound can be
obtained. However, in the case of future supernova neutrino
observations,  one may be able to test this scenario by studying the
neutrino energy spectra.

 SLIM particles should not affect primordial nucleosynthesis. For
masses above $\sim$~MeV, no new light degrees of freedom (dof) are
present at big bang nucleosynthesis epoch. For masses below
$\sim$~ MeV, each scalar would contribute 4/7 dof and each fermion
1 dof. Analysis of cosmological data,  for instance the
combination of the CMB determination of the baryon-to-photon ratio
with primordial light-element abundances observations or with
large scale structure data, lead to an upper bound on the number
of extra dof of $\sim$ 1.5 (at $95\%~$CL)~\cite{BBN1,BBN2,BBN3}.

As far as laboratory constraints  are concerned, light scalar
emission has not been observed in pion and kaon decays. For
kinematically allowed decays a very conservative bound can be
obtained, which constrains the coupling in Eq.(\ref{lagrangian}) to
be $g\lesssim 10^{-2}$~\cite{piondecays1,piondecays2,piondecays3}.
Improving the present experimental bounds seems nevertheless
feasible. For real $\phi$, the upper bound on $m_N$ and the
relatively large value of the coupling (see Eq.~\ref{greal}) promise
observable effects in Kaon and pion decay experiments. This would
make this scenario even more appealing as it could be tested soon.
Many other constraints were discussed in Ref.~\cite{bf} with the
conclusions that this scenario is perfectly viable.

There are certainly many ways to obtain the effective low energy
($SU(2)_L \times U(1)$ breaking) term of Eq.(\ref{lagrangian}) from
an underlying theory. If the particle $N$ is a $SU(2)_L$ singlet,
this interaction term is necessarily effective. It can be obtained,
in particular, from the exchange of an additional scalar
doublet\footnote{A model where such particle content has been
considered in a different mass range can be found in
Ref.\cite{Ma4}.}, an extra vector like fermion singlet, or  an extra
vector-like fermion doublet. The extra particles then have to be
well above the MeV scale. If $N$ is not a $SU(2)_L$ singlet, there
are various possibilities to obtain the same interaction term as in
Eq.(\ref{lagrangian}). This ``fundamental'' Lagrangian was in fact
first proposed in \cite{bf}, with $N$ a mirror fermion (doublet of
$SU(2)_L\times U(1)$). In Ref.~\cite{bf}, $N$ was not a Majorana
particle, so it could not lead to the mechanism described in this
letter. However, one can consider a more sophisticated model where
$N$ is still a mirror particle (it would belong, together with a
charged lepton $E_R$, to a right-handed $SU(2)_L$ doublet) but with
an ``effective'' mass that is induced from $SU(2)_L$ symmetry
breaking. This Majorana mass can be obtained easily from a
``mirror'' seesaw mechanism between $N$ and an extra left-handed
$SU(2)_L$ singlet $N_L$ fermion (to be added to the lagrangian of
Ref.~\cite{bf}). If the mass of the $N_L$ is not far above the
electroweak scale, and if the allowed `$N-N_L-H$' Yukawa coupling is
large enough, this seesaw can lead to a mass $m_N$ well above $\sim
50$ GeV, which can satisfy the constraint on the invisible decay
width of the $Z$-boson. This would require a complex $\phi$, as
explained above. In this model, the lightest $\phi$ component would
be stable for example if, like $N$ and $N_L$, it is odd under a
$Z_2$ symmetry (similarly to models considered in
Ref.~\cite{Ma:1998dn,Ma:2006km}). This solution may be interesting
since, in Ref. \cite{ascasibar}, it is shown that mirror fermions
can lead to the 511 keV line. This interaction term might also
reflect more exotic possibilities. A systematic study of all these
possibilities and related constraints is beyond the scope of this
letter.

\section{Conclusion}

In this letter, we propose a simple scenario, based on a single
interaction term (Eq.~\ref{lagrangian}), where our dark matter
candidate is an electroweak singlet scalar, which interacts with a
Majorana fermion and a left-handed neutrino.

This term generates left-handed neutrino masses through a one-loop
diagram which can be directly related to the SLIM pair annihilation
cross section into neutrinos. This leads to very simple
relationships, Eqs.~\ref{mnu1} and \ref{mnu2}, which constrain the
SLIM particle (our dark matter candidate) to be light. The natural
scale which arises from these equations is the MeV scale (see
Eqs.~\ref{MN} and \ref{Mphisplit}),
 providing a quantitative link between the dark matter
characteristics and the neutrino masses.

A very exciting point is that a dark matter particle with a mass
of only a few MeV also explains the morphology of the 511 keV-line
emission in our galaxy. In addition, for lighter masses, it could
also offer a possible explanation for the LSND signal.

If this picture is experimentally confirmed, our vision of dark
matter in the universe has to be modified. It may play a fundamental
role and even be an active component of the universe (whose presence
is crucial) instead of being a simple relic. Sub-eV neutrino masses
could be the experimental manifestation of MeV particles, possibly
indicating the existence of a low energy theory difficult to access
in collider/accelerator experiments due to the lack of luminosity at
these energies.

Accurate measurements of left-handed neutrino masses (and the study
of neutrino properties in general) could finally open up new
possibilities to answer the question of the origin of the low energy
positrons in our galaxy.

\section*{Acknowledgment}
We would like to thank F.~Boudjema, P. Gagnon, A. Kusenko,  E. Ma,
C. Nicholls, S.~T.~Petcov, M. Raidal, L. Violini for useful
discussions. The authors would like to thank the Theory Division at
CERN for hospitality. SPR is partially supported by the Spanish
grant FPA2005-01678 of the MCT.


\end{document}